\documentclass{scrartcl}
\usepackage[utf8]{inputenc}
\usepackage[T1]{fontenc}
\usepackage{verbatim}
\usepackage{microtype}
\usepackage{mathtools}
\usepackage[english]{babel}
\DeclareMathAlphabet{\mathcal}{OMS}{cmsy}{m}{n}
\usepackage{caption}
\usepackage{subcaption}
\usepackage{tikz}
\usetikzlibrary{matrix}
\usepackage{stmaryrd}
\usepackage{xspace}
\usepackage[hidelinks]{hyperref}
\usepackage{url}
\usepackage[sort,compress,noadjust]{cite}

\title{Challenges for Efficient Query Evaluation\\on Structured Probabilistic
Data}

\author{
\begin{tabular}[t]{c}
Antoine Amarilli \\
{\normalfont Télécom ParisTech, Université Paris-Saclay} \\
{\normalfont antoine.amarilli@telecom-paristech.fr} \\[0.5em]
Silviu Maniu \\
{\normalfont LRI, Université Paris-Sud, Université Paris-Saclay} \\
{\normalfont silviu.maniu@lri.fr} \\[0.5em]
Mikaël Monet \\
{\normalfont Télécom ParisTech, Université Paris-Saclay} \\
{\normalfont antoine.amarilli@telecom-paristech.fr} \\[0.5em]
\end{tabular}
}
\date{}

\usepackage{amssymb}
\usepackage{amsmath}
\usepackage{paralist}
\usepackage{graphicx}
\usepackage{tabularx}
\usepackage{booktabs}

\renewcommand{\leq}{\leqslant}

\renewcommand{\phi}{\varphi}

\makeatletter
\newcommand*{\defeq}{\mathrel{\rlap{\raisebox{0.3ex}{$\m@th\cdot$}}\raisebox{-0.3ex}{$\m@th\cdot$}}=}
\makeatother

\newcommand{\myeat}[1]{}

\newcommand\restr[2]{{
  \kern-\nulldelimiterspace 
  #1 
  _{|#2} 
  }}

\renewcommand{\int}{\mathrm{int}}

\newcommand{\myparagraph}[1]{\paragraph{#1.}}

\begin{document}

\maketitle

\begin{abstract}
  Query answering over probabilistic data is an important task but is generally
  intractable. However, a new approach for this problem has
  recently been proposed, based on \emph{structural decompositions} of input
  databases, following, e.g., tree decompositions.
  This paper presents a vision for a database management system for
  probabilistic data built following this structural approach. We review our existing
  and ongoing work on this topic and highlight many theoretical and practical
  challenges that remain to be addressed.
\end{abstract}

\section{Introduction}
\label{sec:intro}
To have an accurate description of the real world, it is often necessary to
associate probabilities to our observations.
For instance, experimental and scientific data
may be inherently uncertain, because, e.g., of imperfect
sensor precision,
harmful interferences, or 
incorrect modelling.
Even when crisp data can be obtained,
it can
still be the case that we do not trust who retrieved it or how it came to us.
The notion of \emph{probabilistic databases} has been introduced to
capture this uncertainty, reason over it, and query it:
these databases are augmented with
probability information to describe how uncertain each data item is.
Given a probabilistic database $D$ and a query $q$, the \emph{probabilistic
query evaluation problem} (PQE) asks for the probability that the query $q$ holds on
$D$.  Unfortunately, even on the simplest probabilistic database models,
PQE is generally intractable~\cite{dalvi2007efficient}.

One possibility to work around this intractability is to use approximate
approaches, such as Monte Carlo sampling on the data instances. A
different direction was recently explored in~\cite{amarilli2015provenance},
namely, restricting the kind of \emph{input instances} that we allow, in what we
call the \emph{structural approach}.
It is shown in~\cite{amarilli2015provenance}
that the data complexity of PQE is linear if the
instances have \emph{bounded treewidth}, i.e., they can be
structurally decomposed in a tree-like structure where each node must contain
at most $k$ elements, for a fixed parameter $k$. Moreover, 
in~\cite{amarilli2016tractable}, it is shown that bounding the instance treewidth
is \emph{necessary} to ensure the tractability of PQE, because some queries are hard
on \emph{any} unbounded-treewidth instance family (under some conditions).
Hence, bounded-treewidth methods seem to be the right way to make
PQE tractable by the structural approach.

These theoretical works, however, left open the question of practical
applicability: many challenges must still be addressed to implement
a practical system using these techniques.
First, obtaining an
optimal decomposition of an arbitrary instance is \textsc{NP}-hard
\cite{arnborg1987complexity}.
Second, the complexity is only polynomial in the \emph{data}, with the query and
parameter being fixed; this hides a constant which can be exponential in the
width~$k$ and
non-elementary in the query~$q$. Third, we do not know which real datasets
can indeed be decomposed, at least partially, with a small~$k$.

This paper thus presents our vision of a database management system based on
the structural approach, and gives an overview of the
research directions, both theoretical and practical, which we intend to address
to this end.

\section{Probabilistic Query Evaluation: A Structural Approach}
\label{sec:review}
\begin{figure}[t]
\includegraphics[width=\textwidth]{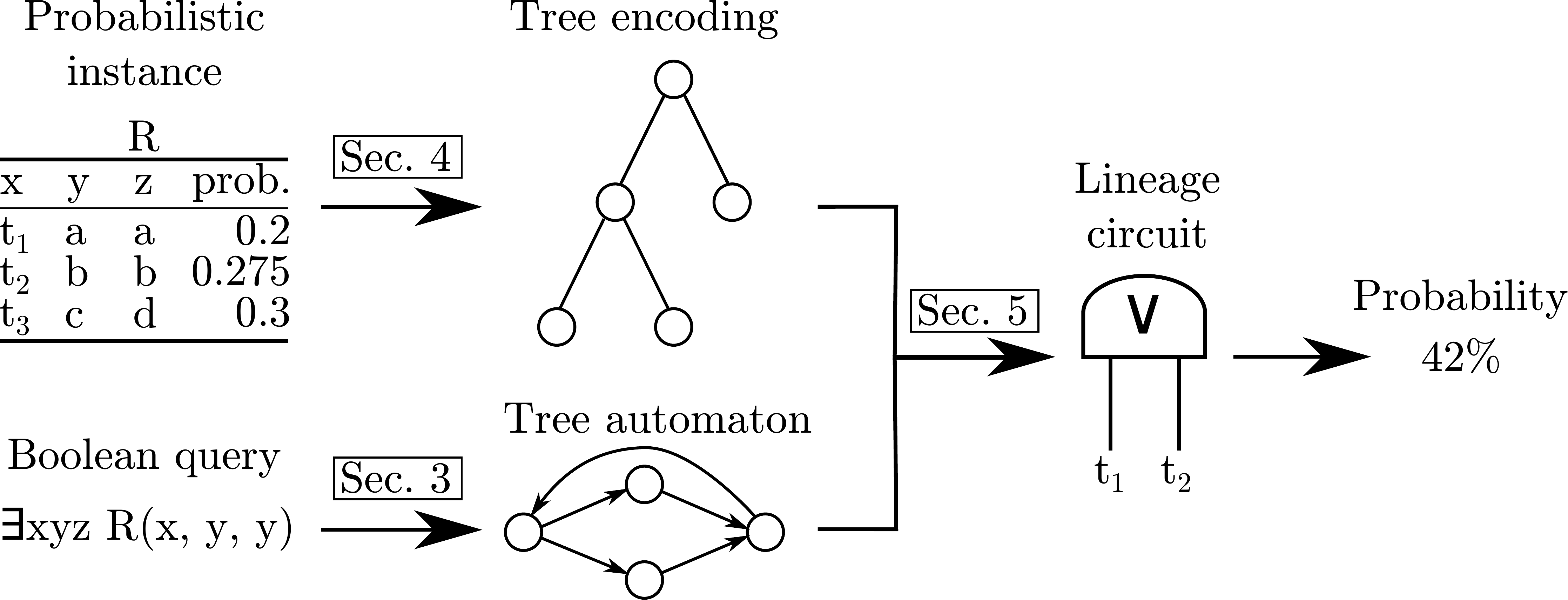}
\caption{Overview of the structural approach for PQE}
\label{fig:courcelle}
\end{figure}

We first review our \emph{structural} approach~\cite{amarilli2015provenance} for probabilistic query
evaluation (PQE).
The approach is illustrated in Figure~\ref{fig:courcelle}.

The approach applies to \emph{tuple-independent (TID) instances} (but
generalizes to more expressive models \cite{thesis}).
Formally, a TID instance $I$ is a relational database~$D$ where each
tuple $t \in D$
has some probability $p_t$. The TID $I$ represents a
probability distribution over the subinstances $D' \subseteq D$ (subsets of
facts): following the independence assumption, the probability of $D'$ is $\prod_{t \in D'} p_t \times
\prod_{t \in D\backslash D'} (1 - p_t)$.

We study the \emph{probabilistic query evaluation} (PQE)
problem: given a \emph{Boolean query} $q$ and TID instance $I$, determine the
probability that $q$ holds on~$I$, i.e., the total probability of the
subinstances of~$I$ that satisfy~$q$.
We refer to the \emph{combined complexity} of PQE when $I$ and $q$ are given as
input; we refer to \emph{data complexity} when $I$ is the input and $q$ is fixed.

The first step of the structural approach
(Section~\ref{sec:automata}) is to translate the query $q$ to a
formalism that can be efficiently evaluated. In the approach
of~\cite{amarilli2015provenance}, following~\cite{courcelle1990monadic}, the query is compiled to a \emph{tree
automaton}, i.e., a finite-state automaton over trees \cite{tata}.
The approach works for expressive queries written in \emph{monadic second-order
logic},
which covers in particular first-order logic and (unions of) conjunctive queries.
This translation of the query is independent from the instance, so does
not affect data complexity; however, it depends on a \emph{parameter} $k$ of the
instance, to be defined soon. 
Intuitively, the automaton represents an algorithm to evaluate the
query on suitable instances.

The second step (Section~\ref{sec:treedecs}) applies to the instance $I$, and computes a \emph{structural
decomposition} of it.
In \cite{amarilli2015provenance}, we compute a \emph{tree decomposition}
\cite{bodlaender2010upper,bodlaender2011lower}, equivalent to junction trees in
graphical models~\cite{lauritzen1988local},
and then a \emph{tree encoding} over a finite
alphabet: the results of \cite{amarilli2016tractable} show that tree
decompositions are
essentially the only possible way to make PQE tractable. The
\emph{parameter} $k$ measures how well $I$ could be decomposed: in our
case, $k$ is the \emph{treewidth}, measuring how close $I$ is to
a tree.
By \emph{treelike} instances, we mean instances whose treewidth is bounded by a
constant.

The third step (Section~\ref{sec:lineages}) is to compute a \emph{lineage} of
the query $q$ on the instance $I$,
i.e., compute an object that represents concisely the subinstances of~$I$ that
satisfy $q$. This object can be used for PQE, as 
what we want to compute is precisely the total probability of this set of
subinstances. Specifically, we compute a
\emph{Boolean lineage circuit} of the tree automaton for the query over the tree
encoding of the instance, according to the construction
of~\cite{amarilli2015provenance}.
This step is purely symbolic and does not
perform any numerical probability computation. 

The fourth and last step is to evaluate efficiently the probability of the query
from this lineage representation, by computing the probability that the circuit
is true. This task cannot be performed
efficiently on arbitrary Boolean circuits, but it is feasible in
our context, for two independent reasons~\cite{amarilli2015provenance,amarilli2016tractable}.
First, this circuit can also be tree decomposed, which allows us to apply a
message-passing algorithm~\cite{lauritzen1988local} for efficient probability
computation. Second, in the case where we made the query automaton
\emph{deterministic}~\cite{tata}, the circuit is actually a d-DNNF~\cite{darwiche2001tractable}, for which probabilistic query evaluation is
tractable.

\section{Efficient Compilation to Expressive Automata}
\label{sec:automata}
Compiling the query to an automaton following the structural approach
of~\cite{amarilli2015provenance}, by applying \cite{courcelle1990monadic},
is generally non-elementary in the query.
This section presents
our main ideas to address this problem: we intend to restrict to \emph{tractable
query fragments}, and to use \emph{more expressive automata targets} to
compile the query more efficiently. These challenges are not specific to PQE;
the next section presents the lineage computation tasks, which are specific to
PQE.

\myparagraph{Efficient Compilation}
Of course, we cannot hope to compile all \emph{Monadic Second Order} logic (MSO)
to automata efficiently, or
even all \emph{conjunctive queries} (CQs).
Indeed, efficient compilation to automata implies that non-probabilistic query
evaluation is also tractable in combined complexity on treelike instances; however,
CQs are already hard to
evaluate in this sense (even on fixed instances).
Hence, we can only hope to compile \emph{restricted} query languages
efficiently.

Many fragments are known from earlier work to enjoy efficient
combined query evaluation. In the database context, for instance,
\emph{acyclic CQs} can be evaluated in polynomial combined
complexity \cite{yannakakis1981algorithms}. This generalizes to
first-order logical sentences that can be written with at most~$k$ variables,
i.e., FO$^k$ \cite{gradel1999logics}. However, it also generalizes to
the \emph{guarded
fragment} (GF) \cite{andreka1998modal}, whose combined complexity is also PTIME, and
where better bounds can be derived if we know the instance treewidth
\cite{berwanger2001games}. The tractability of GF, however, does not capture
other interesting query classes: reachability queries, and more generally 
\emph{two-way regular path queries} (2RPQs) and variants thereof~\cite{barcelo2013querying}, as well as Monadic Datalog as in~\cite{gottlob2010monadic}.

Our first task would thus be to develop an expressive query language that
captures GF, 2RPQs, and Monadic Datalog. Ideally this
fragment should be \emph{parameterized}, i.e., all CQs or all FO queries $q$ could
be expressible in the fragment, up to increasing some parameter $k_q$, with the
compilation being PTIME for fixed~$k_q$ but intractable in~$k_q$. We would then develop an efficient algorithm to
compile such queries to automata that check them on
bounded-treewidth instances, for fixed values of the query parameter $k_q$ and
of the treewidth. Our ongoing work in this direction investigates very recent
extensions of GF with negation and
fixpoints~\cite{benedikt2015complexityc,benedikt2016step}, for which compilation to
automata was studied as a tool for logical satisfiability. We believe that these results,
suitably extended and adapted to query evaluation, can yield to bounded-treewidth
automaton compilation methods that covers the query classes that we mentioned.

\myparagraph{Expressive Automata Targets}
The efficient compilation of queries to automata is made easier by allowing more expressive
automaton classes as the target language. In \cite{amarilli2015provenance}, we
used \emph{bottom-up tree automata}, which process the tree decomposition of the
instance from the leaves to the root. Our idea is to move to more expressive
representations, namely, \emph{two-way alternating automata}~\cite{tata}.  These
automata can navigate through the tree in every direction (including already
visited parts), and thus can be more
concise.
The notion of \emph{alternation} allows automata to
change states based on complex Boolean formulae on the neighboring states, which
also helps for concision.
Indeed, the expressive languages of
\cite{benedikt2016step} are compiled to two-way alternating parity automata,
which further use a parity acceptance condition on infinite runs, to evaluate
fixpoints.

To make automaton compilation more efficient,
another idea is to compile queries to 
automata 
with a concise implicit representation.
In particular, we can use
automata with a \emph{structured} state space: the states are tuples of
Boolean values, and the transition function can be written concisely for each
coordinate of the tuple as a function of the tuples of child states.
It may be possible to capture the tractability of query evaluation for 2RPQs via
automaton methods,
structuring the state space to
memorize separately the regular sublanguages of paths between node pairs.

\section{Obtaining Tree Decompositions}
\label{sec:treedecs}

\myparagraph{Estimating Treewidth}
As we have mentioned, computing the treewidth of an instance directly is an
\textsc{NP}-hard problem. Hence, a practical system using the structural
approach must compute tree decompositions more efficiently, even if this limits
us to non-optimal decompositions.
We intend to experiment with two main kinds of methods to obtain tree decompositions
efficiently: \emph{separator-based} algorithms, which recursively
divide the instance based on various heuristics; and \emph{elimination ordering}
algorithms, where the nodes in the graph are ordered using some measure and
eliminated one by one from the graph~\cite{bodlaender2010upper}. To estimate the
quality of our decompositions, we can also estimate \emph{lower
bounds} on the instance treewidth: for instance
via graph degeneracy
or average degree \cite{bodlaender2011lower}.

\myparagraph{Query-Specific Decompositions}
In some cases, knowledge about the query can help us to obtain better tree
decompositions of the instance. A trivial situation is when we know that the query is only on a
subset of the database relations: we can then ignore the others when
decomposing. More subtly, if we know that specific \emph{joins} are not made by
the query, then we may be able to rewrite the instance accordingly, and lower
the treewidth. For instance, if no $R$- and $S$-atoms share a variable in
the query, then the instance $\{R(a, b), S(b, c)\}$ can be rewritten to $\{R(a,
b), S(b', c)\}$, which may lower the treewidth by disconnecting elements.
We do not
understand this process yet in the general case, but we believe that a theory of
lineage-preserving instance rewritings for a given query (or query class) can be
developed, using the notion of \emph{instance unfoldings} introduced
in~\cite{amarilli2016tractable}.

\section{Tractable Lineage Targets}
\label{sec:lineages}
Once we have compiled the query to an automaton and decomposed the instance to a
tree encoding, our goal is to
compute a \emph{lineage representation} of the automaton on the encoding,
namely, a representation of the subinstances where the query holds, which we
will build as a Boolean circuit.
We can then
use this to perform PQE, by computing the probability of the query as that of the
lineage. In so doing, we need to rely on the fact that the lineage is in a class
of circuits
for which probability can be efficiently computed.

To this end, a first step towards a practical system
is to adapt the methods of \cite{amarilli2015provenance} to the
expressive automaton classes that are needed for efficient query compilation.
We believe that this is possible, but with a twist:
because two-way automata can navigate a tree in every direction, they may go
back from where they came, thus resulting in cyclic runs.
Therefore, it seems that the natural lineages that we would obtain for
alternating two-way automata are \emph{cyclic Boolean circuits}, which we call
\emph{Boolean cycluits}. A semantics for such circuits would need to be defined
based on the semantics of automaton runs and reachable states: we 
believe that the evaluation could follow least fixed-point semantics, and 
be performed in linear time.

Second, we would need to perform efficient probability computation on these
cycluits. One first way to address this would be to eliminate cycles and transform
them to tractable classes of Boolean circuits (e.g., d-DNNFs), which we believe
can be done assuming bounds on the treewidth of the cycluits.
Alternatively, we can
apply message passing methods directly on the cycluits \cite{lauritzen1988local}; 
or we can try to rewrite the automaton to
produce acyclic circuits or even d-DNNFs directly. All these methods would be
generally intractable in the query, which is unsurprising: indeed, PQE is often
intractable even for languages with tractable combined complexity, and efficient
compilation to automata. It would be interesting, however, to identify islands
of tractability; and, in intractable cases, to benchmark the previously
mentioned approaches and see which ones perform best in practice.

Another important direction for a practical system is to be able to evaluate
queries on instances where facts are not independent, i.e., go beyond the TID
formalism. For instance, facts could be present or
absent according to a complex lineage, like the cc-instances of
\cite{thesis}. In this context, new methods can be efficient.
For instance, if the number of probabilistic events is small, performing
\emph{Shannon expansions} on some well-chosen events may make large parts of the
instance deterministic, making the query easier to evaluate on these parts.

\section{Practical Matters}
\label{sec:practical}
We now review possible approaches and directions to implement and evaluate the
structural approach for PQE on real-world datasets.

\myparagraph{Results on Treelike Instances}
In \cite{monet2016probabilistic}, the structural approach has
been compared with one of the existing probabilistic data management systems,
namely MayBMS~\cite{huang2009maybms}.  The instances considered have been
randomly generated to have low treewidth ($\leq 7$).  The results
show that an implementation of the structural approach can perform query
evaluation faster than the exact methods of MayBMS, in cases where there are
many matches and many 
correlations between them.  Indeed, MayBMS does not take
advantage of the fact that the instances are treelike.  However, in this work,
the queries were compiled to automata by hand rather than automatically, and
there was no study of practical datasets.

\myparagraph{Practical Datasets and Partial Decompositions}
A first question is to extend this study to practical datasets, and to
investigate whether such datasets have low treewidth, or whether we can use
approximate decompositions or reasonably low treewidth.
Our preliminary results suggest that some datasets have high treewidth, but
others, in particular transportation networks, have treewidth much smaller than
their size. For instance, the OpenStreetMaps graph of Paris has
over 4 million nodes and 5 million edges, but we estimated its treewidth to be $\leq 521$. We do not
know yet of a theoretical reason explaining why transportation networks generally exhibit this property.

However, this bound is still too large to be practical.
One way to work around this problem is thus to compute a
\emph{partial decomposition} \cite{wei2010tedi} of the instance, i.e., 
a tree decomposition of a part of the instance whose width is at most $k$, with $k$
fixed.
This results in a structure formed
of a forest of instances with treewidth $\leq k$, called the
\emph{tentacles}, that interface with a \emph{core}, i.e., the remaining
facts, whose treewidth is too high and that cannot be decomposed.
Our preliminary experiments have shown that, for some transportation networks, a
partial decomposition for $k = 10$ 
results in a core instance whose size is about 10\% of the original instance.

This decrease in the size of the core, in turn, can potentially have an
immediate effect in the processing of queries. Preliminary results
\cite{maniu2014probtree} have shown that using partial decompositions of fixed
treewidth for probabilistic reachability queries, in conjunction with sampling
in the core graph, can make query processing up to $5$ times faster.

\myparagraph{Tentacle Summarization}
An important problem when computing probabilities on partial decompositions is
the interface between the tentacles and the core, i.e., we must find a way to \emph{summarize}
the tentacles in the core when applying sampling to the core.
As the tentacles are treelike, we can efficiently compute probabilities and
lineages in them: the goal of summarization is to eliminate the tentacles and
replace them by
\emph{summary} facts that are added to the core.
In the case of simple queries, such as reachability queries, the
summary facts can have the same semantics as in the original instance, but this
does not seem to generalize to arbitrary queries: it may even be the case that
some queries cannot be rewritten to the summary facts 
while remaining in the same language.

Having summarized the tentacles, we may also answer
queries approximately via \emph{sampling}, using the (exact)
tentacle summaries added to the core: as the instance is now smaller, this
process can be performed faster.

\bibliographystyle{abbrv}
\bibliography{main,thesis,topic,antoine}

\begin{thebibliography}{10}

\bibitem{thesis}
A.~Amarilli.
\newblock {\em Leveraging the Structure of Uncertain Data}.
\newblock PhD thesis, Télécom ParisTech, 2016.
\newblock 2016-ENST-0021. \url{https://tel.archives-ouvertes.fr/tel-01345836}.

\bibitem{amarilli2015provenance}
A.~Amarilli, P.~Bourhis, and P.~Senellart.
\newblock Provenance circuits for trees and treelike instances.
\newblock In {\em Proc.\ ICALP}, 2015.
\newblock \url{https://arxiv.org/abs/1511.08723}.

\bibitem{amarilli2016tractable}
A.~Amarilli, P.~Bourhis, and P.~Senellart.
\newblock Tractable lineages on treelike instances: Limits and extensions.
\newblock In {\em Proc.\ PODS}, 2016.
\newblock \url{https://arxiv.org/abs/1604.02761}.

\bibitem{andreka1998modal}
H.~Andr{\'e}ka, I.~N{\'e}meti, and J.~van Benthem.
\newblock Modal languages and bounded fragments of predicate logic.
\newblock {\em J. Philosophical Logic}, 27(3), 1998.
\newblock \url{http://doai.io/10.1023/A:1004275029985}.

\bibitem{arnborg1987complexity}
S.~Arnborg, D.~G. Corneil, and A.~Proskurowski.
\newblock Complexity of finding embeddings in a k-tree.
\newblock {\em SIAM Journal on Algebraic and Discrete Methods}, 8(2), 1987.
\newblock
  \url{http://melodi.ee.washington.edu/~bilmes/grg/Arnborg_k_tree_1987.pdf}.

\bibitem{barcelo2013querying}
P.~Barcel{\'o}~Baeza.
\newblock Querying graph databases.
\newblock In {\em Proc.\ {PODS}}, 2013.
\newblock \url{https://users.dcc.uchile.cl/~pbarcelo/pods001i-barcelo.pdf}.

\bibitem{benedikt2016step}
M.~Benedikt, P.~Bourhis, and M.~Vanden~Boom.
\newblock A step up in expressiveness of decidable fixpoint logics.
\newblock In {\em Proc.\ {LICS}}, 2016.
\newblock
  \url{https://www.cs.ox.ac.uk/people/michael.vandenboom/papers/LICS16-gnfpup-long.pdf}.

\bibitem{benedikt2015complexityc}
M.~Benedikt, B.~Ten~Cate, T.~Colcombet, and M.~V. Boom.
\newblock The complexity of boundedness for guarded logics.
\newblock In {\em Proc.\ {LICS}}, 2015.
\newblock
  \url{https://www.cs.ox.ac.uk/people/michael.vandenboom/papers/LICS15-gnfpb-long.pdf}.

\bibitem{berwanger2001games}
D.~Berwanger and E.~Gr{\"a}del.
\newblock Games and model checking for guarded logics.
\newblock In {\em Proc.\ {LPAR}}, 2001.
\newblock \url{http://www.lsv.ens-cachan.fr/Publis/PAPERS/PS/BG-lpar01.ps}.

\bibitem{bodlaender2010upper}
H.~L. Bodlaender and A.~M. C.~A. Koster.
\newblock Treewidth computations {I}. {U}pper bounds.
\newblock {\em Information and Computation}, 208(3), 2010.
\newblock
  \url{http://www.sciencedirect.com/science/article/pii/S0890540109000947}.

\bibitem{bodlaender2011lower}
H.~L. Bodlaender and A.~M. C.~A. Koster.
\newblock Treewidth computations {II}. {L}ower bounds.
\newblock {\em Information and Computation}, 209(7), 2011.
\newblock
  \url{http://www.sciencedirect.com/science/article/pii/S0890540111000836}.

\bibitem{tata}
H.~Comon, M.~Dauchet, R.~Gilleron, C.~L\"oding, F.~Jacquemard, D.~Lugiez,
  S.~Tison, and M.~Tommasi.
\newblock Tree automata: Techniques and applications, 2007.
\newblock \url{https://gforge.inria.fr/frs/download.php/file/10994/tata.pdf}.

\bibitem{courcelle1990monadic}
B.~Courcelle.
\newblock The monadic second-order logic of graphs. {I}. {R}ecognizable sets of
  finite graphs.
\newblock {\em Inf. Comput.}, 85(1), 1990.
\newblock
  \url{http://www.sciencedirect.com/science/article/pii/089054019090043H}.

\bibitem{dalvi2007efficient}
N.~Dalvi and D.~Suciu.
\newblock Efficient query evaluation on probabilistic databases.
\newblock {\em VLDBJ}, 16(4), 2007.
\newblock \url{https://homes.cs.washington.edu/~suciu/vldbj-probdb.pdf}.

\bibitem{darwiche2001tractable}
A.~Darwiche.
\newblock On the tractable counting of theory models and its application to
  truth maintenance and belief revision.
\newblock {\em J. Applied Non-Class.\ Log.}, 11(1-2), 2001.
\newblock \url{https://arxiv.org/abs/cs/0003044}.

\bibitem{gottlob2010monadic}
G.~Gottlob, R.~Pichler, and F.~Wei.
\newblock Monadic {D}atalog over finite structures of bounded treewidth.
\newblock {\em TOCL}, 2010.
\newblock \url{https://arxiv.org/abs/0809.3140}.

\bibitem{gradel1999logics}
E.~Gr{\"a}del and M.~Otto.
\newblock On logics with two variables.
\newblock {\em TCS}, 224(1), 1999.
\newblock
  \url{http://www.sciencedirect.com/science/article/pii/S0304397598003089}.

\bibitem{huang2009maybms}
J.~Huang, L.~Antova, C.~Koch, and D.~Olteanu.
\newblock {MayBMS}: a probabilistic database management system.
\newblock In {\em Proc.\ {SIGMOD}}, 2009.
\newblock \url{http://maybms.sourceforge.net/download/sigmod2009_demo.pdf}.

\bibitem{lauritzen1988local}
S.~L. Lauritzen and D.~J. Spiegelhalter.
\newblock Local computations with probabilities on graphical structures and
  their application to expert systems.
\newblock {\em JRSS Ser.\ B}, 1988.
\newblock
  \url{https://www.eecis.udel.edu/~shatkay/Course/papers/Lauritzen1988.pdf}.

\bibitem{maniu2014probtree}
S.~Maniu, R.~Cheng, and P.~Senellart.
\newblock {ProbTree}: A query-efficient representation of probabilistic graphs.
\newblock In {\em Proc.\ {BUDA}}, 2014.
\newblock \url{http://pierre.senellart.com/publications/maniu2014probtree.pdf}.

\bibitem{monet2016probabilistic}
M.~Monet.
\newblock Probabilistic evaluation of expressive queries on bounded-treewidth
  instances.
\newblock In {\em Proc.\ PhD Symposium of SIGMOD/PODS}. ACM, 2016.
\newblock \url{https://zenodo.org/record/58133/}.

\bibitem{wei2010tedi}
F.~Wei.
\newblock {TEDI}: Efficient shortest path query answering on graphs.
\newblock In {\em Proc.\ SIGMOD}, 2010.
\newblock \url{http://doai.io/10.1145/1807167.1807181}.

\bibitem{yannakakis1981algorithms}
M.~Yannakakis.
\newblock Algorithms for acyclic database schemes.
\newblock In {\em Proc.\ {VLDB}}, 1981.
\newblock \url{http://doai.io/10.1145/320083.320091}.

\end{thebibliography}

\end{document}